\documentstyle[aps,twocolumn,epsf]{revtex}

\newcommand{\be}{\begin{eqnarray}}
\newcommand{\ee}{\end{eqnarray}}

\renewcommand{\vec}{\bbox}

\begin{document}

\title{
Fermion mass-function in thermal  $\tau_{3}-QED3$  }

\author{Georg Triantaphyllou}

\address{
 Institut f\"ur Theoretische Physik, Technische 
Universit\"at M\"unchen\\
 James-Franck-Strasse, D-85748 Garching, Germany }

\date{\today}

\maketitle

\begin{abstract}
The gap equation for fermions in a version of 
thermal  $QED$  in three dimensions 
is studied numerically in the  
Schwinger-Dyson formalism.  The interest in this theory 
has been recently revived since it has been proposed as a 
model of high-temperature superconductors.
We include 
wave-function renormalization in our equations and use a non-bare vertex. 
We then have to solve a system of two integral equations 
by a relaxation algorithm.   
Fermion and photon self-energies varying independently with
energy and momentum are used, which should produce 
more accurate results than in the previous literature. 
The behaviour of
the theory with increasing temperature and number of
fermion flavours is then carefully analyzed. 
\end{abstract}

\narrowtext

\section{THE PROBLEM}
The study of $QED$ in three dimensions during the past years
has revealed very interesting 
physics in this theory \cite{bunch}-\cite{volo}.    
At finite temperature, a version of $QED$, i.e.
$\tau_{3}-QED$, has been proposed as a model of  
 high-temperature 
superconductivity \cite{Mavro}. The  behaviour of the theory
with increasing temperature $T$ and number of fermion flavours $N_{f}$ was
recently studied numerically 
in the Schwinger-Dyson formalism in a bare-vertex approximation \cite{georg}. 
This approximation corresponds to a particular truncation of the 
Schwinger-Dyson hierarchy,  
and its effects, along with wave-function  
renormalization effects, could be important \cite{Penni}, 
especially in connection with the behaviour of the theory when  
the number of fermion flavours is varied. 
This has led to several interesting studies at
zero-temperature trying to include these effects \cite{Nash}-\cite{Maris}, in
order to solve the controversy on whether there is a critical 
number of fermion flavours beyond which there is no mass generation. At 
finite temperature, similar studies have so far used several approximations
that we will relax in this study. 

The first purpose of this work is to provide an accurate 
value  for the ratio $r=2M(0,0)/k_{B}T_{c}$ \cite{Mavro}, where
$M(0,0)$ is the fermion mass function at zero momentum and energy
and $T_{c}$ is the critical temperature beyond which there is no
mass generation.  Apart from being theoretically very interesting, 
this quantity is of direct physical relevance since it can be
compared to corresponding values measured in certain
high-temperature superconductors. The second  purpose   
is to provide 
a reliable phase diagram of the theory with respect to $T$ and $N_{f}$. 
We continue to consider momentum- and energy-dependent self-energies, but
we include wave-function renormalization effects and a non-bare vertex.  

This complicates the study considerably and constitutes a non-trivial
step forward, since instead of having only
one gap equation, a system of two  integral equations has 
to be solved. We continue
to neglect the imaginary parts of the photon polarization
functions and of the fermion self-energy for simplicity, 
as other studies have done so far \cite{instant}-\cite{ian2}. 
Their inclusion, even though in principle necessary,
would double the number of coupled equations to be solved,
which would render the numerical study too complicated for our
algorithm and the computer power at hand.

\section{THE EQUATIONS} 
We will employ the Schwinger-Dyson formalism in order to gain
information on the momentum-dependent fermion self-energy and the
non-perturbative physics behind it.   
The Schwinger-Dyson equation for the fermion self-energy is given by
\begin{equation}
S^{-1}(p) = S_{0}^{-1}(p) 
- e^{2}\int \frac{d^{3}k}{(2\pi)^{3}}
\gamma^{\mu}S(q)\Delta_{\mu\nu}(k)\Gamma^{\nu}(p,q) 
\end{equation}

\noindent 
where $q = p - k$, $e$ is the dimensionful gauge coupling of the theory 
which we will take to be constant throughout this study, 
$\Delta_{\mu\nu}$ is the photon propagator with $\mu, \nu = 0, 1, 2$, 
$\Gamma^{\nu}$ is the
full photon-fermion vertex, $\gamma^{\mu}$ is a four-dimensional
representation of the $\gamma$-matrices,
$S_{0}$ is the bare fermion propagator, and the
 finite-temperature
fermion propagator in the real-time formalism 
is given by 
\begin{eqnarray} 
S(p)& =& \left(\left(1+A(p)\right)p\!\!\!/ + \Sigma(p)\right)
\times \left( 
 \frac{1}{\left(1+A(p)\right)^{2}p^{2} + \Sigma^{2}(p)}\right. \nonumber \\
 & &-
 \left. \frac{2\pi\delta((1+A(p))^{2}p^{2}+\Sigma^{2}(p))}
 {e^{\beta|p_{0}|}+1}\right),  
 \end{eqnarray}  

\noindent where $\beta = 1/k_{B}T$,  
$A(p)$ is the wave-function renormalization function,  
$\delta$ is the usual Dirac function and we have made a rotation to
Euclidean space.
Note that we avoid the matrix form that the propagator has in 
this formalism, 
since the Schwinger-Dyson equation that we have written down  
involves only a one-loop diagram directly,   
so complications due to the field-doubling problem
do not arise \cite{Land}. 
However, it has to be noted that a more careful treatment involving 
the imaginary parts of the self-energies would involve the full matrix
propagators \cite{Smilga}. The resulting equations would in that case be
unfortunately  too 
complicated to be solved even numerically, without serious truncations.
Moreover, due to the 
broken Lorenz invariance at finite temperature, 
the wave function renormalization
could in principle affect differently the $p_{0}$ and $|\vec{p}|$
propagator components, i.e.
we should replace $(1+A(p))p\!\!\!/$ by 
$((1+A(p))\gamma^{0}+a)p_{0} + (1+B(p))\gamma^{i}p_{i}$ 
with $i = 1, 2$. For simplicity we will restrict ourselves
to situations where $a=0$, which correspond to a zero
chemical potential, and we will work in the 
approximation where $A(p) = B(p)$ also for non-zero 
temperatures as done in similar studies \cite{ian2}. 

For the vertex  $\Gamma^{\nu}(p,q)$ we use the ansatz 
$\Gamma^{\nu}(p,q) = \left(1+A(q)\right)\gamma^{\nu}$, where
$A(q)$ is the same wave-function renormalization function
appearing in the fermion propagator. 
It is unfortunately not symmetric in the vertex momenta, but its
use simplifies the numerical algorithm considerably.
Even though
this vertex does not satisfy {\it a priori}
the Ward-Takahashi identities, it is expected to incorporate the basic
qualitative features of a non-perturbative vertex at zero temperature 
when used in a Schwinger-Dyson context \cite{Miranski}.  
It has also been used in a  
finite-temperature case \cite{ian2}, supported by the qualitative 
agreement of the results of this ansatz with the ones obtained by more
elaborate treatments \cite{Penni}. Furthermore, in the problem of the
present study it gives results
similar to a symmetrized vertex \cite{georg2}. 

Moreover, the photon propagator in the Landau gauge
is given by \cite{Mavro} 
\begin{equation}
\Delta_{\mu\nu}(k) = \frac{Q_{\mu\nu}}{k^{2}+\Pi_{L}(k)}
+\frac{P_{\mu\nu}}{k^{2}+\Pi_{T}(k)}
\end{equation} 
\noindent where 
\begin{eqnarray}
Q_{\mu\nu}&=&(\delta_{\mu 0}-k_{\mu}k_{0}/k^{2})\frac{k^{2}}{\vec{k}^2}
(\delta_{\nu 0} - k_{\nu}k_{0}/k^{2}) \nonumber \\ 
P_{\mu\nu}&=&\delta_{\mu i}(\delta_{ij} - 
k_{i}k_{j}/\vec{k}^2)\delta_{\nu j}
\end{eqnarray}

\noindent  with $i, j = 1, 2$, 
and where we neglect its temperature-dependent delta-function
part since it is expected to give a vanishingly small contribution
\cite{georg},  \cite{ian1}, \cite{delref}.
The longitudinal and 
transverse photon polarization functions $\Pi_{L}$ and $\Pi_{T}$
are given explicitly in \cite{georg}  and taken from \cite{ian1}, 
where they are calculated in a massless-fermion approximation and
where wave-function renormalization and  vertex effects cancel
only for specific vertex choices \cite{Maris}. 
One should in principle couple the expressions for $\Pi_{L,T}$
in our system of integral equations in a self-consistent manner, but
unfortunately this would render our numerical algorithm too 
complicated.

Identifying the parts of this equation with the same spinor structure, 
we reduce the problem to that of a system of two three-dimensional  
integral equations involving two functions varying independently with 
$p_{0}$ and $|\vec{p}|$.
The equations take the following form:
\begin{eqnarray}
 &&M(p_{0},|\vec{p}|)=\frac{\alpha}{N_{f}(1+
 A(p_{0},|\vec{p}|))}\int 
\frac{dk_{0}|\vec{k}|d|\vec{k}|d\theta}{(2\pi)^{3}}  
\times \nonumber \\&&\nonumber \\ 
&& \times 
\frac{ M(q_{0},|\vec{q}|)}{q^{2}+ 
M^{2}(q_{0},|\vec{q}|)} 
\sum_{P=L,T} 
\frac{1}{k^{2}+\Pi_{P}(k_{0},|\vec{k}|)}  
\nonumber \\ && \nonumber \\ 
&&-\frac{\alpha}{N_{f}(1+ A(p_{0},|\vec{p}|))} 
\int\frac{|\vec{k}|d|\vec{k}|d\theta}{(2\pi)^{2}}  
\frac{M(E,|\vec{q}|)}
{2E(e^{\beta E} + 1)} 
 \times  \nonumber \\&&\nonumber \\&&\times
 \sum_{\epsilon=1,-1}\sum_{P=L,T} 
\frac{1}{(p_{0}-\epsilon E)^{2}+\vec{k}^{2}+
\Pi_{P}(p_{0}-\epsilon E,|\vec{k}|)}  \nonumber \\&&\nonumber \\ 
&&A(p_{0},|\vec{p}|)=\frac{\alpha}{N_{f}p^{2}}\int 
\frac{dk_{0}|\vec{k}|d|\vec{k}|d\theta}{(2\pi)^{3}}  
\frac{1}{q^{2}+ 
 M^{2}(q_{0},|\vec{q}|)} 
\times \nonumber \\&&\nonumber \\  
&&\times \left(\frac{Q(p_{0},\vec{p},k_{0},\vec{k})}{k^{2}+
\Pi_{L}(k_{0},|\vec{k}|)} 
+\frac{P(p_{0},\vec{p},k_{0},\vec{k})}{k^{2}
+\Pi_{T}(k_{0},|\vec{k}|)} \right) 
 \nonumber \\&&\nonumber \\ 
&&-  \frac{\alpha}{N_{f}p^{2}} 
\int\frac{|\vec{k}|d|\vec{k}|d\theta}{(2\pi)^{2}}  
 \frac{1}
{2E(e^{\beta E} + 1)} 
 \times  \nonumber \\&&\nonumber \\&&\times \sum_{\epsilon=1,-1}\left( 
\frac{Q(p_{0},\vec{p},p_{0}-\epsilon E,\vec{k})}{(p_{0}-\epsilon E)^{2}
+\vec{k}^{2}+
\Pi_{L}(p_{0}-\epsilon E,\vec{k})} + \right. \nonumber \\&&\nonumber\\ 
&&+\left.
\frac{P(p_{0},\vec{p},p_{0}-\epsilon E,\vec{k})}{(p_{0}-\epsilon E)^{2}
+\vec{k}^{2}+
\Pi_{T}(p_{0}-\epsilon E,\vec{k})}\right), 
\label{eq:fingap}
\end{eqnarray}

\noindent 
where $\alpha = e^{2}N_{f}$, 
and  it is more convenient to work with the mass function 
$M(p_{0},|\vec{p}|)=
\Sigma(p_{0},|\vec{p}|)/(1+ A(p_{0},|\vec{p}|))$. 
We also sum over the photon polarizations $P=L, T$ and
over the two roots of the delta function by introducing $\epsilon=1,-1$. 
The quantity $E$ is  approximated by the relation 
$E^{2} \approx |\vec{q}|^{2} + M^{2}(0,0)$ \cite{georg}, where
use of the delta-function property $\delta(ax)=\delta(x)/|a|$ has been
made.

\noindent  Furthermore, the functions $Q$ and $P$ are given by
\begin{eqnarray}
Q(p_{0},\vec{p},k_{0},\vec{k})&=& 2\left(p_{0}-\frac{(pk)k_{0}}{k^{2}}\right)
\frac{k^{2}}{\vec{k}^2}\left(q_{0}-\frac{(qk)k_{0}}{k^{2}}\right)
\nonumber\\
P(p_{0},\vec{p},k_{0},\vec{k})&=& 2\left(\vec{p}\;\vec{q}
-\frac{(\vec{p}\vec{k})(\vec{k}\vec{q})}{\vec{k}^2} \right) -pq. 
\end{eqnarray}
\noindent  One can easily check that 
$Q + P = - 2(pk)(kq)/k^{2}$, which would reproduce the result of 
\cite{ian2} if one takes $\Pi_{L}(k)=\Pi_{T}(k)=\Pi(k)$ and 
switches to imaginary-time formalism. 

After inspecting the equations, we note that the vertex ansatz we 
chose makes the integral giving the function $A(p_{0},|\vec{p}|)$
depend only on the function $M$ and 
independent of $A$. On the other hand, the equation for
$M(p_{0},|\vec{p}|)$ has to be solved self-consistently, and it
actually always accepts, apart from the solutions we will seek,
the trivial solution as well. An analytical study of such a system
would not be possible without severe approximations, so we proceed to
the numerical solution of the equations given above.

\section{NUMERICAL RESULTS}
The numerical method used to solve the system of equations
is the same as the one presented in  \cite{georg}.  The physical
quantities of interest do not vary substantially with our 
ultra-violet (UV) cut-off, since at the effective UV cut-off
$\alpha$ of the theory they are already negligibly small \cite{georg2}. 
The solution at $T=0$ and $N_{f}=2$ for the functions
$\Sigma(p_{0},|\vec{p}|)$ and 
$ - A(p_{0},|\vec{p}|)$ is given in 
Figs. 1 and 2 respectively. The general variation of 
these functions with momentum does not change with increasing temperature
or varying of $N_{f}$, even though the overall scale of $\Sigma$
drops fast with $N_{f}$. 
The function $\Sigma(p_{0},|\vec{p}|)$ falls 
as expected with increasing momentum, and is of the same form
as the mass function $M(p_{0},|\vec{p}|)$. 
The function $A(p_{0},|\vec{p}|)$ is always in the range
between -1 and 0 as required \cite{Claude}, it is tending to zero
for increasing momenta, and it is of the same
form and magnitude as the approximate form used in \cite{ian2}. 

For a given number of fermion flavours $N_{f}$, when the temperature
exceeds some critical value there is no solution for the fermion mass function 
but the trivial one. 

\begin{figure}\epsfxsize=3cm
\centerline{\epsfbox{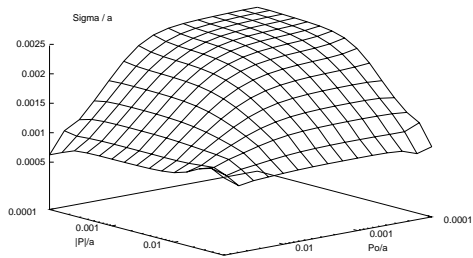}}
\vskip4mm
\caption{The fermion self-energy $\Sigma(p_{0},|\vec{p}|)$ 
 at zero temperature and for $N_{f} = 2$,
$\Lambda_{UV}/\alpha = 0.1$, as a function of energy and momentum in
logarithmic scale. All quantities are scaled by $\alpha$.}  
\label{fig1}
\end{figure}
\begin{figure}\epsfxsize=3cm
\centerline{\epsfbox{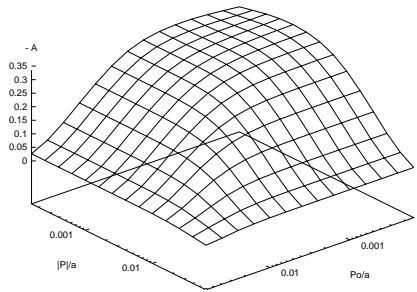}}
\vskip4mm
\caption{The opposite of the  
wave-function renormalization $-A(p_{0},|\vec{p}|)$ at zero temperature
and for $N_{f} = 2$, $\Lambda_{UV}/\alpha = 0.1$.}  
\label{fig2}
\end{figure}
\begin{figure}\epsfxsize=3cm
\centerline{\epsfbox{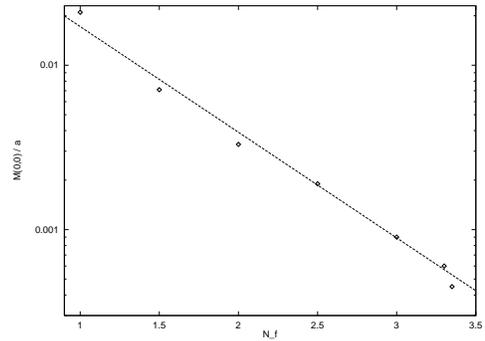}}
\vskip4mm
\caption{The fermion mass function at zero momentum and zero temperature, 
scaled by $\alpha$ and on a logarithmic scale,  
with respect to
$N_{f}$ for a ratio $\Lambda_{UV}/\alpha = 0.1$. 
We fit our results with the curve 
$M(0,0)/\alpha = e^{-1.48N_{f}}/13.25$.   
 Values of $N_{f}$ larger than 3.35 are not
considered, because then the self-energy falls below the IR-cut-off.}  
\label{fig3}
\end{figure}
\begin{figure}\epsfxsize=3cm
\centerline{\epsfbox{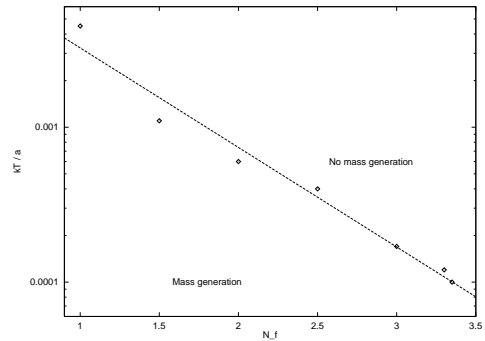}}
\vskip4mm
\caption{The  
phase diagram of the theory with respect to the temperature scaled by 
$\alpha$ and in
logarithmic scale and
the number of fermion flavours, for a ratio
$\Lambda_{UV}/\alpha = 0.1$. We fit the data points with the
curve $k_{B}T/\alpha = e^{-1.48 N_{f}}/70$. This curve should not
be extrapolated for $N_{f} ~^{>}_{\sim}\; 3.35$.}  
\label{fig4}
\end{figure}

The value of the ratio $r$ is
concentrated approximately around $r \approx 10.6$, which is 
comparable to the values obtained in \cite{georg}  
which neglected $A(p_{0},|\vec{p}|)$.
The ratio $r$ found is also comparable to  
typical values obtained in Ref. \cite{ian2}, which includes wave-function
renormalization effects but uses several approximations which we 
were able to by-pass in this study. We confirm therefore that adding 
these effects does not influence the behaviour of the theory in a
significant way.
We have  to note moreover
that our $r$ values are somewhat larger than
the value $r \approx 8$ measured for
some high-temperature superconductors \cite{Schles}.
However, we could be overestimating this ratio 
because of a possibly poor convergence of the algorithm for
temperatures close to the critical one.  

At zero-temperature, this theory is known to exhibit also an interesting
behaviour with the number of fermions $N_{f}$. 
In Fig. 3 we plot the zero-momentum and zero-temperature fermion
mass function with respect to  $N_{f}$. We fit the data with the
exponential curve $e^{-1.48N_{f}}/13.25$. 

At $N_{f} \approx 3.35$, the mass function is still roughly four 
times larger than the cut-off. 
When $N_{f} ~^{>}_{\sim}\; 3.35$,  
our algorithm  does not converge and the mass function 
tends fast below  the IR cut-off. 
This behaviour
could indicate that $N_{f} \approx 3.35$  is some critical point beyond
which dynamical mass generation is impossible. 

The value of $N_{f}$ we find is  remarkably close to the one
quoted in the numerical study of \cite{Maris}, and it is also close to
our previous result \cite{georg}.
A  similar study \cite{Nash}, 
which includes a calculation of the
fermion-field anomalous dimension to 
second-order in $1/N_{f}$, 
predicts a critical  value 
$N_{f}  \approx 3.28$, which 
is only slightly larger than the 
theoretical prediction neglecting wave-function 
renormalization which gives 
 $N_{f} = \frac{32}{\pi^{2}} \approx 3.24$ \cite{appel}, and  
 quite close to the value  we find numerically. 

In Fig. 4 we plot the phase diagram of the theory with respect to 
$N_{f}$ and $k_{B}T$. It separates two regions of the parameter space
which either allow or do not allow dynamical mass generation.  
The choice of an exponential fitting curve was only made to describe
``phenomenologically"
the general tendency of the data and to provide a measure  for a
$r$-ratio independently of $N_{f}$, and is reminiscent of the results
in Ref. \cite{Penni} but with a somewhat steeper slope. 
However, there are also studies that predict a
non-analytic behaviour of $\Sigma$ for $N_{f}$ near its critical
value \cite{appel}. 
Lack of convergence of the algorithm and
fall of the mass function below the IR cut-off  however do not allow us to
test the precise behaviour of the theory near $N_{f} \approx 3.35$, and
it is also clear that the fitting curves should not be extrapolated for 
$N_{f}$ larger than this value. 

\section{DISCUSSION}
We were able to  
solve a system of two coupled integral equations for the
fermion mass function and wave-function renormalization for
 a finite-temperature version of three-dimensional $QED$,   
by applying a numerical relaxation technique.  
One main result is a $r$-ratio of about 10.6, which 
is close to previous numerical studies, confirming that including 
wave-function renormalization, in conjunction with a particular
non-bare fermion-photon vertex,  does not affect the theory in a
significant way.
The other important result is the
existence of a possibly critical fermion 
flavour number of roughly 3.35, which
is  also consistent with some theoretical expectations and 
other numerical results. 

It is the first time
that such a study includes the momentum and energy dependence of the 
fermion and photon self-energies and of wave-function 
renormalization in a gap equation with a non-bare vertex,  
and this allows a more reliable description of the behaviour of the theory. 
We estimate the numerical uncertainty for 
the values quoted, which comes mainly from the
convergence criteria imposed, at about $\pm 10\%$. 
The next step for future studies should be 
the inclusion of the imaginary parts of the self-energies in the 
equations and the relaxing of the approximation $A(p)=B(p)$, 
which could in principle influence the results of this
investigation.

\acknowledgments
I thank N. Mavromatos, P. Henning and  A. Smilga
for useful discussions.
This research is supported by an {\it Alexander von Humboldt Fellowship}.

\end{document}